\documentstyle[12pt,epsf,epsfig,wrapfig]{article}
\begin{document}
\begin{center}
{\bfseries Gluonic effects in vector meson
photoproduction at large momentum transfers}
\vskip 1mm
N.I. Kochelev$^{1,2\alpha}$ and
V. Vento$^{3,\beta}$.
\vskip 1mm
{\small
(1) {\it
Bogoliubov Laboratory of Theoretical Physics, JINR, Dubna, Russia} \\
(2) {\it Institute of Physics and Technology, Almaty, 480082,
Kazakhstan}\\
(3) {\it Department de F$\acute i$sica Te$\grave o$rica-IFIC, Universitat
de Val$\grave e$ncia-CSIC, Spain}
\\
E-mail~~{\it $^\alpha$:~kochelev@thsun1.jinr.ru,~~
$^\beta$:~vicente.vento@uv.es,~~}}
\end{center}
\vskip 1mm
\begin{abstract}
Non-perturbative QCD mechanisms are of fundamental
importance in strong  interaction physics. In particular,
the flavor singlet axial anomaly leads  to a gluonic pole
mechanism which has been shown to explain the
$\eta^{\prime}$ mass, violations of the OZI rule and  more
recently  the proton spin. We show here that the interaction
derived from the gluonic pole exchange  explains the high
momentum transfer behavior of the photoproduction
cross sections of vector  mesons  at JLab energies.
\end{abstract}

\section{Introduction}

Recently the CLAS Collaboration at JLab  has published new data on $\rho$- and
$\phi$-meson photoproduction at large momentum transfers \cite{CLAS1, CLAS2}.
The main goal of these data is to provide understanding for the mechanism
responsible for the violation of the OZI rule in strong interactions
\cite{titov,williams,zhao}.

The usual approach to the description of exclusive reactions consists  in the
use of some effective non-perturbative reaction model at small momentum
transfers and  hard perturbative gluonic exchange mechanisms  at large momentum
transfers \cite{laget1}. These  approaches are unable to   explain the large
spin effects at large $-t$ in  exclusive reactions \cite{krish}.
One  example is the
 large $A_{nn}$ asymmetry in proton-proton elastic
scattering which  can be understood only if some helicity violating
contribution is present at such momentum transfers. The pQCD
exchanges in these
approaches lead only to quark helicity conserving contributions and therefore
can not be the dominant mechanisms for exclusive reactions in the few GeV
region.

We have proposed a non-perturbative mechanism \cite{kochvento} for exclusive
reactions at large momentum transfers, whose relevance we have studied in
$\phi$ meson electromagnetic production  at large momentum transfer,  which we
here generalize to  $\rho$- and $\omega$-meson photoproduction off the nucleon.
The new ingredient of our  model is the gluonic contribution arising from the
QCD mechanism associated with the flavor singlet axial anomaly and describable
in terms of an additional gluonic pole in the amplitudes. The gluonic pole was
introduced in \cite{veneziano1} to describe  features of OZI violation in the
pseudoscalar meson nonet. It has been shown to provide a natural explanation of
proton spin \cite{veneziano2}. Recently, applications to
radiative decays of pseudoscalar mesons, the muon anomalous magnetic moment and
the determination of the photon structure $g_1^{\gamma}$ function in polarized
deep inelastic scattering have been discussed \cite{shore}. The origin for this
pole is the periodicity of QCD the  potential as a  function of the topological
charge and the existence of instantons tunneling between the various classical
vacua of the theory \cite{diakonov}.

By showing the relevance of the gluonic pole exchange in vector meson
photoproduction at large momentum transfers, we unveil a non-perturbative QCD
effect, which is neither contained in the low energy effective theories nor in
the pQCD ingredients of previous calculations.

\section{Vector meson photoproduction off the nucleon}

It is very well known that at large energy and small  momentum transfer the
main contribution to the photoproduction of the vector mesons comes from the
pomeron exchange. In the low energy region one can expect the dominance of the
pomeron at small $-t$  only for $\phi$ meson electromagnetic
photoproduction due
to OZI rule. For $\rho$ and $\omega$-meson production the contribution from
secondary Reggeon exchanges might be important. In the case of photoproduction
the leading Regge trajectory with appropriate  quantum
numbers is the $f_2$ meson
trajectory.  Subleading Regge exchanges, which might  give a significant
contribution,  are $\pi$-meson exchange  for $\rho$- and $\omega$-meson
production and $\eta$-meson exchange for $\phi$-meson production \cite{titov}
(Fig.1a).

\begin{figure}
\begin{center}
\epsfig{file=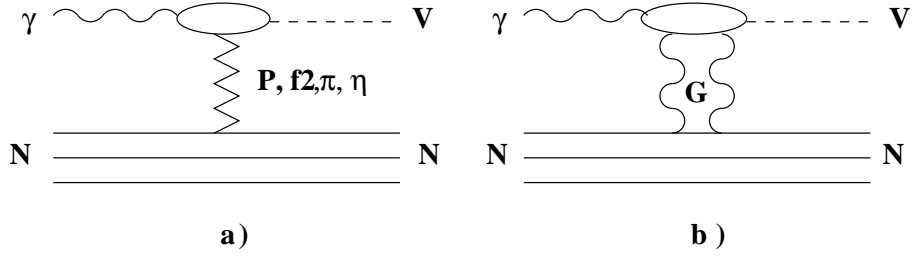,width=12cm}
\caption {{\small
a) Contributions from the pomeron, $f_2$, $\pi$ ($\eta$)
to $\rho$  and $\omega$ ($\phi$) photoproduction;
b) Contribution from the $G$-pole exchange to
 vector-meson photoproduction.}}
\label{vm11}
\medskip
\end{center}
\end{figure}

We use the Donnachie-Landshoff (DL) model \cite{DL1} to
describe the soft  pomeron contribution to the vector meson production
differential cross section,

\begin{equation}
\frac{d\sigma_P}{dt}= \frac{81 m_V^3 \beta_0^2\beta_q^2
\Gamma_{e^+e^-}^V}
{\pi \alpha_{\rm em}}
\frac{F(t)^2\mu_0^4}{(2\mu_0^2 + Q^2 + m_V^2 - t)^2(Q^2 + m_V^2 - t)^2}
 \left( \frac{S}{S_0} \right)^{2 \alpha_P(t) - 2},
\label{cross3}
\end{equation}

\noindent where

\begin{equation}
F(t)=\frac{4M_N^2-2.8t}{(4M_N^2-t)(1-t/0.7)^2}
\label{form1}
\end{equation}
is the electromagnetic nucleon form factor. The pomeron-quark couplings
$\beta_0=2$ GeV$^{-1}$, $\beta_s=1.5$ GeV$^{-1}$
are obtained  from a fit to the  total $pp$,
$\pi p$ and  $Kp$ cross sections \cite{DL2},
 $\mu_0^2=1.1 GeV^2$,
$S_0=4 GeV^2$ and the pomeron trajectory is
$\alpha_P(t)=\alpha_P(0)+\alpha_P^\prime t$, with
$\alpha_P(0)=1.08$ and
$\alpha_P^\prime=0.25$ GeV$^{-1} $.
The effect of the
$f_2$ trajectory can be taken into account by multiplying the equation for the
pomeron  contribution Eq.(\ref{cross3}) by the factor
\begin{equation}
F_{P+f_2}=1+
2A(S,t)cos\left(\frac{\pi}{2}(\alpha_{f_2}(t) - a_P(t))\right)+
A(S,t)^2
\label{f2}
\end{equation}

\noindent with

\begin{equation}
A(S,t)=\frac{\beta_{f_2}^2}{\beta_0^2}\left(\frac{S}{S_1}
\right)^{\alpha_{f_2}(t) - 1}
\left(\frac{S}{S_0}\right)^{1- \alpha_P(t)},\nonumber
\end{equation}
where
 the $f_2$ trajectory is
$\alpha_{f_2}(t)=0.55+0.9 t$, and the coupling to quarks $\beta_{f_2}=
4.32$ was taken also from the DL fit to the total hadron cross sections
\cite{DL2}.

The contribution of the pseudoscalar exchange to the cross section can be
calculated by using the same procedure as in the DL approach to the pomeron
contribution \cite{DL1} \footnote{We do not include the interaction of the
$\eta$-meson with the $u-$ and $d-$ quarks because its contribution to
$\rho$ and $\omega$-meson production is much smaller than the contribution
of the $\pi$-meson.}

\begin{figure}
\begin{center}
\epsfig{file=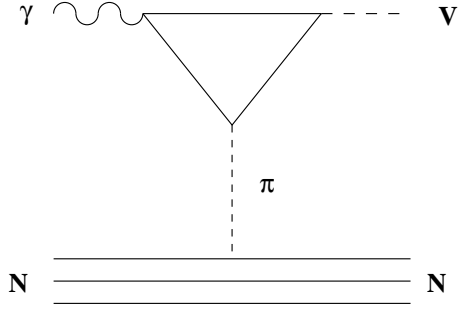,width=6.cm}
\caption {{\small
Contribution of the triangle graph to the $\pi$ exchange amplitude.}}
\label{triangle}
\medskip
\end{center}
\end{figure}
Direct calculation of the triangle diagram in Fig.2
\footnote{We do not use the conventional approach for the
calculation of the pseudoscalar contribution which consists in using
as input  the experimental values of widths for $V\rightarrow \gamma \pi^0
(\eta)$ \cite{titov} since one has large uncertainties in the estimates
 due to the poor knowledge  of  the $V \gamma \pi^0 (\eta)$ form-factors.}
at the quark level with the quark-meson interactions
\begin{eqnarray}
L_{\pi qq}&=&\frac{g_{\pi^0qq}}{2m_q}(\bar u\gamma_\mu\gamma_5u-
\bar d\gamma_\mu\gamma_5d)\partial^\mu\pi^0,\nonumber\\
L_{\eta ss}&=&\frac{g_{\eta ss}}{2m_s}
\bar s\gamma_\mu\gamma_5s\partial^\mu\eta
\label{PS1}
\end{eqnarray}
gives the result
\begin{equation}
\frac{d\sigma_{PS}}{dt}= -\frac{K_{PSV} m_V g_{PSNN}^4
\Gamma_{e^+e^-}^Vt}
{\pi \alpha_{\rm em}}
\frac{FPS(t)^2}{(S-M_N^2)^2(t - M_{PS}^2)^2},
\label{PS}
\end{equation}
where $K_{\pi\rho}=3/100$, $K_{\pi\omega}=9K_{\pi\rho}$,
$K_{\eta\phi}=3$ and we have used the constituent quark
relations $2m_q=m_V$, $g_{\pi^0qq}=3/5g_{\pi^0NN}$ and
$g_{\eta ss}=2g_{\eta NN}$.

For numerical estimates the value $g_{\pi NN}=13.28$ \cite{pion}
for pion-nucleon and the $SU(3)_f$ value $g_{\eta NN}=3.52$
for $\eta$-nucleon coupling \cite{titov} have been used.
In Eq.(\ref{PS}) the pseudoscalar-nucleon form-factor is taken
of monopole form
\begin{equation}
FPS(t)=\frac{\Lambda^2-M_{PS}^2}{\Lambda^2-t}
\label{formps}
\end{equation}
with $\Lambda=0.8 GeV$ \cite{strikman}.
The Reggeization of the pseudoscalar meson exchange can be perform
in the standard way (see for example \cite{laget2}).
It is done by substituting the pseudoscalar propagator in the following way,
\begin{equation}
\frac{1}{t-M_{PS}^2}\rightarrow
\left(\frac{S}{S_1}\right)^{\alpha_{PS}(t)}
\frac{\pi\alpha^\prime}{sin\pi\alpha_{PS}(t)}
\frac{1+e^{-i\pi\alpha_{PS}(t)}}{2\Gamma(1+\alpha_{PS}(t))}.
\label{regge}
\end{equation}
The pseudoscalar trajectories are taken as
\begin{equation}
\alpha_{PS}(t)=\alpha^\prime(t-M_{PS}^2)
\label{traj}
\end{equation}
with slope $\alpha^\prime=0.9 GeV^{-2}$.

It can be shown that the effect of Reggeization leads to the
multiplication of
Eq.(\ref{PS}) by the Regge factor
\begin{equation}
R=\left(\frac{S}{S_1}\right)^{2\alpha_{PS}(t)}\Gamma(1-\alpha_{PS}(t))
cos^2(\frac{\pi\alpha_{PS}(t)}{2}).
\label{regge2}
\end{equation}

The pomeron contribution to vector meson photoproduction is
shown in Figs. 3, 4 and 5 by  long-dashed lines.  It underestimates
the low $-t$ cross-section for $\rho$ and $\omega$
production at JLab energies but  describes rather well the
low $-t$ $\phi$-meson production. The total contribution of
the DL pomeron and $f_2$ exchange  to the cross section
for  $\rho$ and $\omega$  photoproduction  is
presented in Figs. 3 and 4 by dashed-dot lines. The contribution of
the pion exchange in Figs. 3 and 4 and $\eta$-meson
exchange in Fig. 5 is shown by short-dashed line.  It is
apparent that pseudoscalar exchange  is important  only for
$\omega$ production at very small momentum transfers.
Our result is in contradiction with the result of ref.
\cite{titov} where a nonreggeized pion exchange
contribution to vector meson production was calculated.
In our approach the suppression of the pseudoscalar
contribution at large $-t$ is not very sensitive to the
$-t$ dependence of the $\pi(\eta) NN$ and $\pi (\eta) \gamma V$
form-factors and the absolute value  of the  pseudoscalar
coupling with quarks. This suppression is due to the
smallness of the Regge factor $(S/S_1)^{2\alpha_{PS}(t)}$
in the cross-section for large $-t$.  The effect of the
Reggeization  for the case of the
$\pi$ contribution to $\rho$ production is shown in Fig. 6.
This effect
is very large and can not be  neglected. The pomeron and
Regge  contributions  describe the data for all mesons
rather well in the small $-t\leq 0.5$ GeV$^2$ region.
However,  for large $-t\geq 1$ GeV$^2$,  all these
contributions  deviate strongly from the data.

\begin{figure}
\begin{center}
\epsfig{file=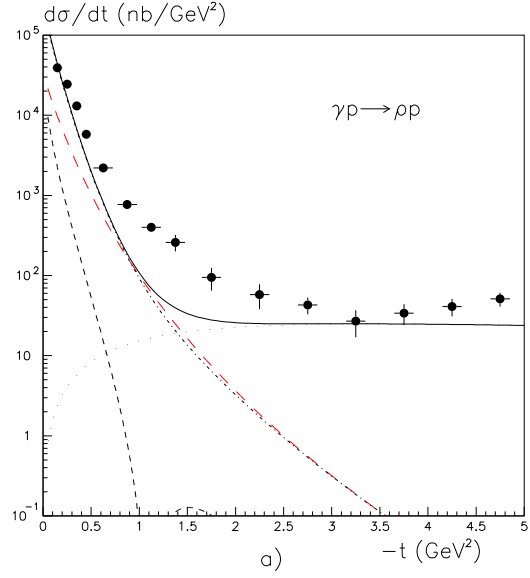,width=8.cm}
\caption {{\small
The contributions of the  pomeron (long-dashed line), the sum of pomeron and
$f_2$ exchanges (dashed-dot line), pion (short-dashed line)   and
$G$- exchange (dotted line)  to
$\rho$-meson production at $E_\gamma=3.82 GeV$.
The solid line represents the total contribution.
The data are from the CLAS
Collaboration \cite{CLAS1}.}}
\label{vm1}
\medskip
\end{center}
\end{figure}
\begin{figure}
\begin{center}
\epsfig{file=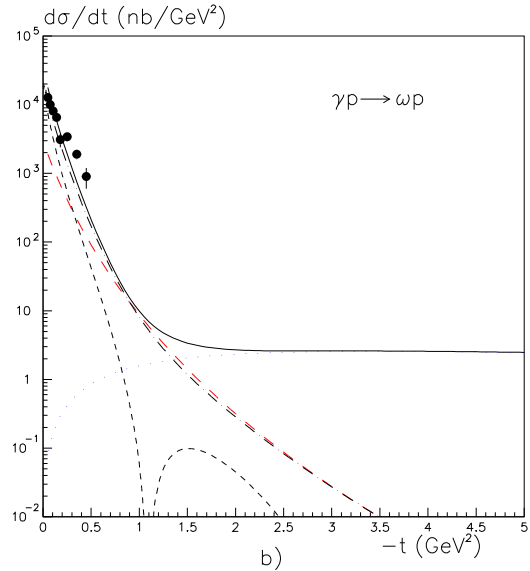,width=8.cm}
\caption {{\small
The contributions of the  pomeron , the sum of pomeron and
$f_2$ exchanges, pion
and $G$-exchange to
$\omega$-meson production at $E_\gamma=3.87 GeV$.
The notation follows that of the previous figure and the data
are from \cite{barber}.}}
\medskip
\end{center}
\end{figure}
\begin{figure}
\begin{center}
\epsfig{file=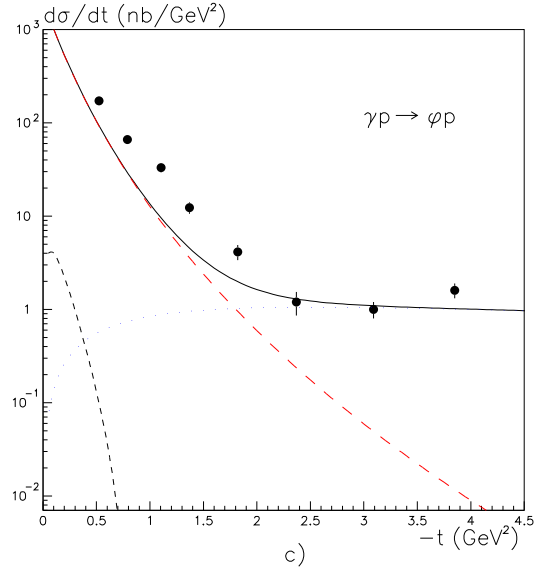,width=8.cm}
\caption {{\small
The contributions of the  pomeron, the sum of pomeron and
$f_2$ exchanges, $\eta$ (short-dashed line)
 and
$G$-exchange  to the  $\phi$-meson
photoproduction at $E_\gamma=3.6 GeV$.
The notation for the pomeron and $f_2$ contributions
follows that of the previous figures and the data
are from \cite{CLAS2}.
}}
\medskip
\end{center}
\end{figure}
\begin{figure}
\begin{center}
\epsfig{file=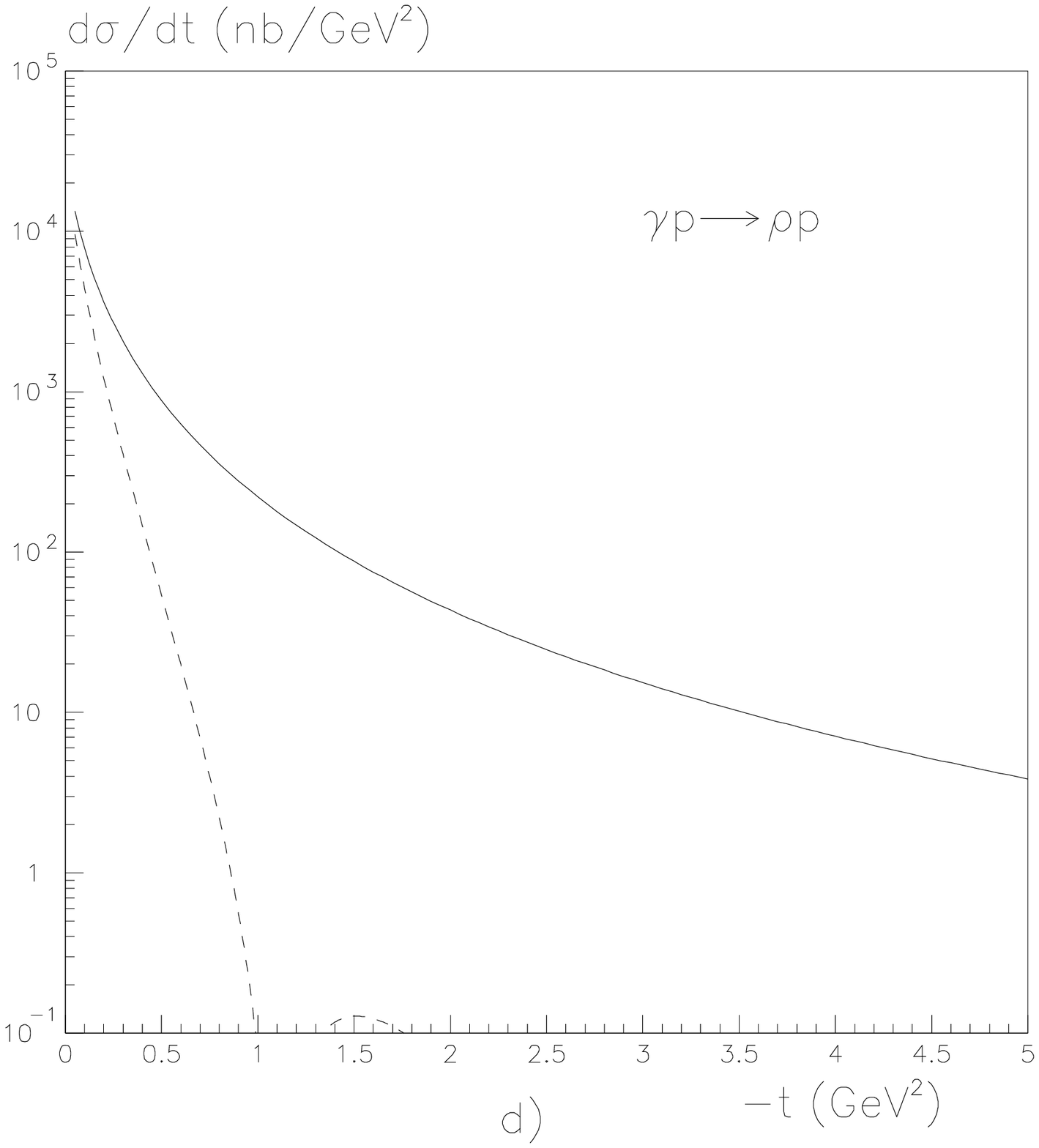,width=8.cm}
\caption{{\small The effect of $\pi$ reggeization in $\rho$ production
is shown. }}
\medskip
\end{center}
\end{figure}

\section{Anomalous gluonic exchange}

In our previous work \cite{kochvento} a new gluonic,
so-called  $G$-pole, exchange  was introduced to explain
the large $-t$ electromagnetic $\phi$ meson production at
JLab.  This exchange, not related to the exchange of any mesonic
state,  arises from the nontrivial topological structure
of the QCD vacuum. In some sense this exchange represents
the interaction of the quarks in the photon with the quarks
in the proton generated by the excitations of the QCD
vacuum.

In QCD  the anomalous gluonic axial current
\begin{equation}
K_{\mu}=\frac{\alpha_s}{4\pi}\epsilon_{\mu\nu\alpha\beta}A_\nu^a
(\partial_\alpha
A_\beta^a+\frac{g_s}{3}f_{abc}A_\alpha^bA_\beta^c)
\label{current}
\end{equation}
has a massless pole
\begin{equation}
{i\int d^4x
e^{ikx}<0|TK_{\mu}(x)K_{\nu}(0)|0>_{k\rightarrow0}}\rightarrow
\frac{g_{\mu\nu}}{k^2}\lambda^4
\label{correlator}
\end{equation}
which is related to the topological susceptibility of the
QCD vacuum
\begin{equation}
\chi(0)=-\lambda^4=i\int d^4x<0|TQ(x)Q(0)|0>,
\label{correlator2}
\end{equation}
where $Q(x)=\partial_\mu K_\mu $ is the density of topological
charge. This pole has been called the gluonic ghost
pole \cite{veneziano1} in the literature and $G$-pole by us
\cite{kochvento}.

The diagonalization of the propagator matrix for the ghost
operator $Q$ and  quark nonet states $\pi^0,\eta^8$ and $\eta^\prime$
leads to the following  propagator
\begin{equation}
<GG>=-A
\label{ghost}
\end{equation}
which we called $G$-propagator.
$A$  in Eq.(\ref{ghost}) is determined by
the topological susceptibility of QCD vacuum \cite{shore}
\begin{equation}
A\approx -\chi(0)\vert_{YM}
=\frac{f_\pi^2}{6}
(m_{\eta^\prime}^2+m_\eta^2-2m_K^2)\approx (180 MeV)^4.
\label{top}
\end{equation}

We should stress that the $G$-propagator Eq.(\ref{ghost}) does
not depend on the momentum transfer $k^2$ and therefore the
effective interaction induced by the ghost exchange is {\it
point-like}. This property is responsible for the large
$G$-pole  contribution to  vector meson photoproduction at
large momentum transfers as will be shown next.

Let us estimate the contribution of $G$-pole exchange to vector meson
photoproduction (Fig.1b).
To calculate this diagram the values of couplings $g_{GNN}$ and
$g_{G\gamma V}$ are needed.
We will use the fact that
the contribution of the $G$-pole to the physical
amplitudes in the flavor singlet channel leads to results which are
different from the
predictions of the OZI rule. This mechanism produces, for
the isosinglet axial-vector nucleon form factor at
zero momentum transfer, the following generalized $U(1)$
Golberger-Treiman relation \cite{veneziano3}
\begin{equation}
2M_NG_A^0=Fg_{\eta^\prime NN}+2N_fAg_{GNN}=F_{\eta_0}g_{\eta_{0NN}},
\label{GT}
\end{equation}
where $F\approx\sqrt{2N_f}f_\pi$, $f_\pi=93$ MeV and
 $g_{GNN}$ is  $G$-nucleon coupling constant.
The experimental value of
$G_A^0 \approx 0.3$ extracted from experimental data on nucleon
spin-dependent structure function $g_1(x,Q^2) $ \cite{anselmino},
 allows us to estimate the value of $G$-nucleon
coupling constant as
\begin{equation}
g_{GNN}\approx-\frac{0.3M_N}{N_fA}\approx-89.35 GeV^{-3}.
\label{gGNN}
\end{equation}

The $G$-pole contribution to the $\eta^\prime(\eta)\rightarrow\gamma\gamma$
decay has been discussed in \cite{veneziano3} where
a modified formula  for the  effective coupling of the $\eta^\prime$ meson
with photons has been obtained,
\begin{equation}
Fg_{\eta^\prime\gamma\gamma}+2N_fAg_{G\gamma\gamma}=
\frac{4}{\pi}\alpha_{em}.
\label{Ggg}
\end{equation}

Let us consider the $G$-pole contribution to the $\eta^\prime\gamma V$
coupling, where $V$ stands for $\rho^0$, $\omega$ and
$\phi$.
The interaction of the vector mesons with
the quarks is assumed to be photon-like
\begin{equation}
L_{\rho qq}= C_{\rho}(\bar u\gamma_\mu u-\bar d\gamma_\mu d)\rho_\mu, {\ }
L_{\omega qq}= C_{\omega}(\bar u\gamma_\mu u+\bar d\gamma_\mu d)\omega_\mu, {\ }
 L_{\phi ss}=C_\phi\bar s\gamma_\mu s\phi_\mu .
\label{phiss}
\end{equation}
Due to this vector meson-photon analogy the generalization
of Eq.(\ref{Ggg})
to the case of the $\eta^\prime\rightarrow\gamma V$
amplitude is straightforward
\begin{equation}
Fg_{\eta^\prime\gamma V}+2N_fAg_{G\gamma V}=
k_VC_V\sqrt{\frac{\alpha_{em}}{\pi^3}},
\label{GgPhi}
\end{equation}
where $k_\phi=-1$, $k_\rho=3$ and $k_\omega=1$.
The values of $g_{\eta^\prime\gamma\phi}$, $g_{\eta^\prime\gamma\omega}$  and
 $g_{\eta^\prime\gamma\rho}$
 can be
extracted from the experimental widths
$\Gamma_{\phi\rightarrow\eta^\prime\gamma}$,
 $\Gamma_{\omega\rightarrow\eta^\prime\gamma}$
and
$\Gamma_{\eta^\prime\rightarrow\rho\gamma}$.

The vector meson couplings with
quarks  are not well known. In order to estimate them  we use
a constituent quark model  calculation for the
$\rho$ and $\omega$ couplings
\begin{equation}
C_\rho=C_\omega=\frac{g_{\omega NN}}{3}\approx 3.45,
\label{VDM}
\end{equation}
where $g_{\omega NN}=10.35$ is taken  from  \cite{titov}
and
for the
$\phi$ coupling we use the NJL model prediction \cite{volkov}
\begin{equation}
C_\phi=-5.33.
\label{NQM}
\end{equation}
Our final estimate for the $G$-vector meson couplings is
$|g_{G\gamma\rho}|=11.02$ GeV$^{-4}$, $|g_{G\gamma\omega}|=3.57$
$GeV^{-4}$
  and
$|g_{G\gamma\phi}|=1.91$  $GeV^{-4}$.

We are now ready to calculate $G$-pole contribution to
vector meson photoproduction.
The absolute value of the $G$-contribution  strongly
depends on its couplings with the nucleon, photon and vector meson
and is given by
\begin{equation}
\frac{d\sigma_{G}}{dt} = -\frac{A^2g_{G\gamma\phi}^2g_{GNN}^2
t(t-M_\phi^2)^2}{64\pi(S-M_N^2)^2}F_1(t)^2,
\label{gcross}
\end{equation}
where
$F_1(t)=1/(1-t/M_{f_1}^2)^2$
is the flavor singlet axial form factor of the nucleon
with the value of $M_{f_1}$   equals to  the mass of the flavor singlet
$f_1(1285)$-meson \cite{kochvento}.

It should be noted that the $G$-exchange induces a nucleon
spin-flip, which  produces a factor $t$ in
Eq.(\ref{gcross}), and leads to an  additional enhancement
of the $G$-pole contribution at large $-t$, with respect to
the pomeron contribution, which is nonspin-flip
Eq.(\ref{cross3}).  Moreover the energy dependence of the
$G$  contribution corresponds to that of a fixed pole with zero
Regge slope.  Therefore the large $t$ Regge suppression
in vector meson production
given by $({S}/{M_N^2})^{2\alpha_R^\prime t}$ with the
slope  $\alpha_R^\prime\approx 0.9$ GeV$^{-2}$ for the
usual Regge trajectories, e.g.
$\pi^0$ and $\eta $, is
absent for the $G$-exchange.

The result of the calculation of the $G$-pole  contribution
  is presented
in Fig. 3, 4 and 5  by the dotted lines. The sum of the
pomeron, $f_2$, $\pi^0$, $\eta$  and $G$-pole contributions
 are shown by the solid lines.
It is evident that $G$-exchange determines the behavior
of the cross section at large $-t$ and our model
based only on pomeron, secondary Regge exchange and $G$-pole
 contributions reproduces
the main features of a data.
The deviation from data at intermediate values of $-t$ might be
related to additional contributions from the pomeron and Regge cuts
or to non-linear $t$ dependences of Pomeron and $f_2$
trajectories\cite{khoze}.

\section{Conclusion}

The gluonic degrees of freedom  play a very important  role
in the  $\rho$- $\omega$- and $\phi$-meson photoproduction
at JLab energies. At small $-t$ the cross-section is
described rather well by the Donnachie-Landshoff soft
pomeron  and secondary Regge exchanges. At large $-t$ an
extremely interesting phenomena, related to the complex
structure of QCD vacuum, takes place. We have shown that at
large momentum transfers, the point-like interaction
induced by the  $G$- pole exchange, related to ghost pole
in the correlator of the anomalous gluonic axial currents,
gives the dominant contribution. We should stress that
vector meson photoproduction is only one of the possible
exclusive processes where the $G$-pole contributes. We plan to
extend our considerations here to the study of other reactions at
large momentum transfers.

As has been mentioned the G-pole exchange leads to
a nucleon helicity flip and therefore can be separated from
the helicity conserving pQCD hard gluonic exchanges
in processes with polarized particles.

Recently we have shown that at large energy and large
momentum transfer a new anomalous trajectory with very
large intercept $\alpha_{f1}\approx 1$ and very small slope
$\alpha^\prime\approx 0 $ called $f_1$ is needed to explain
elastic $pp$, $p\bar p$ and vector meson production at HERA
\cite{f1}. It would be interesting to find a relation
between this $f_1$ trajectory and the $G-$ exchange
similarly as was discussed recently for the relation between
pomeron  and S-channel multi-glueball $0^{++}$ exchange
\cite{shuryak}.

We are grateful to S.B.Gerasimov, W.-D.Nowak  and V.L.Yudichev for
useful discussion. We would like to thank M. Battaglieri
and R. Schumacher for providing us with the CLAS data. This
work was partially  supported by DGICYT PB97-1227,
RFBR-01-02-16431, INTAS-00-00366 grants and by the
Heisenberg-Landau program.

\end{document}